\documentclass{emulateapj}
\newcommand{\kms}{km~s$^{-1}$\,}
\newcommand{\msun}{${\cal M}_\odot$\,}
\newcommand{\masyr}{mas~yr$^{-1}$}

\begin{document}

% for float placement:
\renewcommand{\topfraction}{1.0}
\renewcommand{\bottomfraction}{1.0}
\renewcommand{\textfraction}{0.0}

\title{Orbits of Five Triple Stars}
%\altaffilmark{\dag}}

\author{Andrei Tokovinin}
\affil{Cerro Tololo Inter-American Observatory | NSF's NOIRLab, Casilla 603, La Serena, Chile}
\email{atokovinin@ctio.noao.edu}

\author{David W. Latham}
\affil{Center for Astrophysics | Harvard \& Smithsonian, 60 Garden Street, Cambridge, MA 02138, USA}
\email{dlatham@cfa.harvard.edu}, 

\begin{abstract}
Joint analysis of radial velocities and position measurements of five
hierarchical stellar systems is undertaken to determine elements of their inner
and outer orbits and, whenever possible, their mutual inclinations. The
inner and outer periods are 12.9 and 345 yr for HD 12376 (ADS 1613),
1.14 and $\sim$1500 yr for HD 19971 (ADS 2390), 8.3 and 475 yr for HD
89795 (ADS 7338),  1.11 and 40 yr for HD 152027,  0.69 and 7.4 yr for HD
190412. The latter system with its co-planar and quasi-circular orbits belongs
to the family of compact planetary-like hierarchies, while the orbits in
HD 12376 have mutual inclination of 131\degr. 
\end{abstract} 


%-------------------------------------------------------------
\section{Introduction}
\label{sec:intro}

Observational data on hierarchical  stellar systems with three or more
components are still incomplete  and fragmentary because discovery and
study  of   these  objects  is  challenging.  Large   range  of  their
separations and periods calls  for the use of complementary techniques
and/or long-term  monitoring.  Here we  determine spectroscopic orbits
of  inner  subsystems in  five  known  visual  binaries in  the  solar
neighborhood. In two  cases we  detect astrometric  signal of the
inner  orbit and  therefore determine  mutual orbit  orientation.  Our
work contributes information on  the architecture of hierarchical systems
related to their origin and encoded in the mass ratios, period ratios,
eccentricities, and mutual inclinations.

\begin{deluxetable*}{ccc cc lc lc   }
\tabletypesize{\scriptsize}
%\tablenum{7}
\tablewidth{0pt}
\tablecaption{List of multiple systems
\label{tab:list} }
\tablehead{
%\multicolumn{3}{c}{Discoverer} &
\colhead{WDS} &
\colhead{HD} & 
\colhead{HIP} & 
\colhead{$V$} &
\colhead{$\varpi$} & 
\colhead{Outer} & 
\colhead{$P_{\rm out}$} &
\colhead{Inner} & 
\colhead{$P_{\rm in}$ } \\
 &        &  &
\colhead{(mag)} & 
\colhead{(mas)} & 
& \colhead{(yr)} &
& \colhead{(yr)} 
}
\startdata
02022$+$3643 & 12376 & 9500    & 8.14  & 19.9: & V  & 345   & V,S1 & 12.9  \\ 
03122$+$3713 & 19771 & 14886   & 7.88  & 20.58 & V  & 1500: & S1   & 1.14   \\
10217$-$0946 & 89795 & 50747   & 8.01  & 15.88 & V  & 475   & A,S1 & 8.39  \\  
16446$+$7145 & 152027 & 81961  & 8.66  & 14.85 & V,S1 & 39.7 & S1  & 1.11   \\
20048$+$0109 & 190412 & 98878  & 7.69  & 20.6: & V,S1 & 7.4 & S1,A & 0.69   
\enddata
\tablecomments{Explanation of columns: (1) WDS code \citep{WDS};
(2) HD number; 
(3) Hipparcos number;
(4) visual magnitude; 
(5) parallax (colons mark less certain parallaxes);
(6) type of outer orbit;
(7) outer period;
(8) type of inner orbit;
(9) inner period.
 }
%\tablecomment{Explanation of columns...}
\end{deluxetable*}

\begin{figure}
\epsscale{1.1}
\plotone{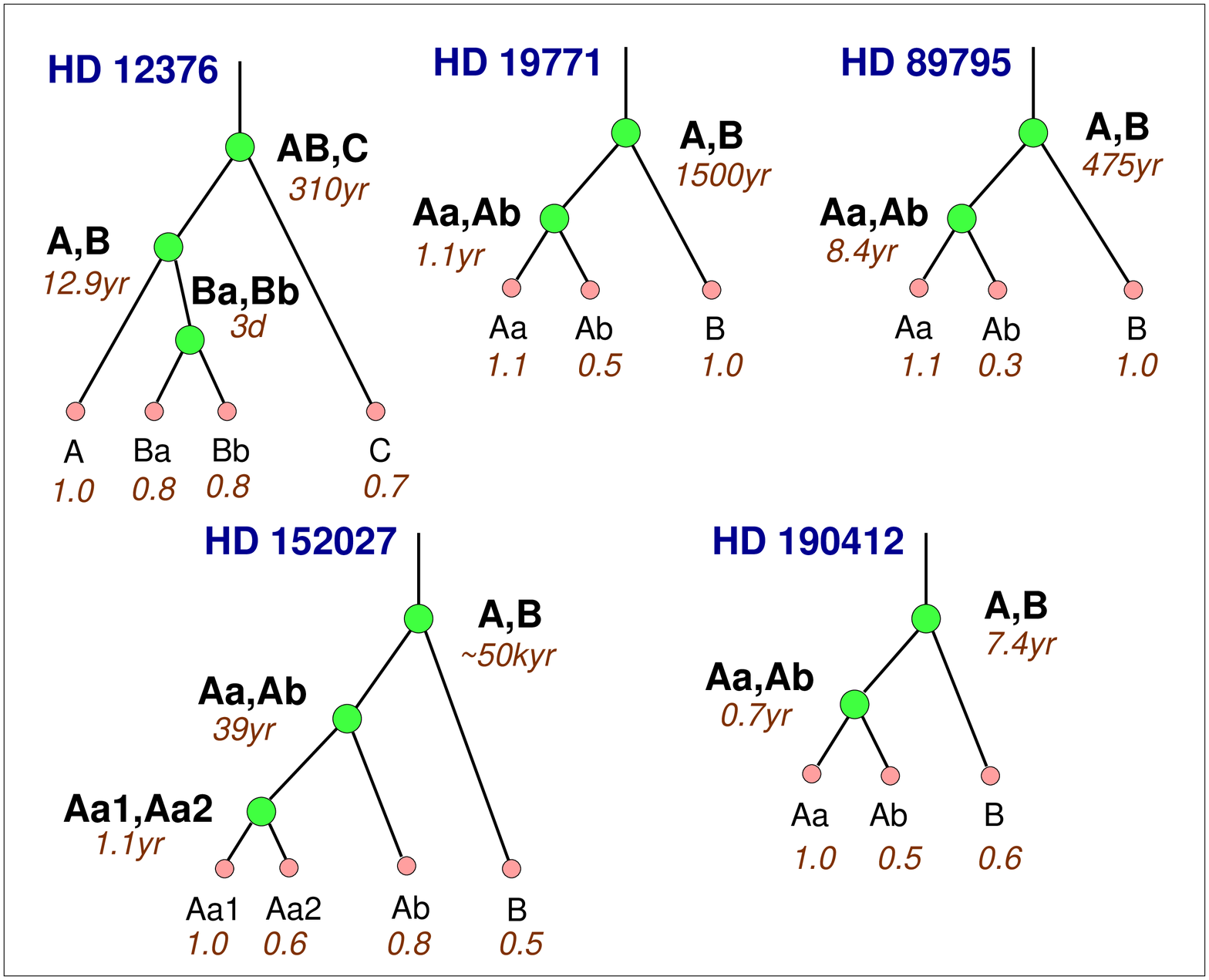}
\caption{Structure of multiple systems. Green circles refer to pairs,
  smaller pink circles to components. Numbers in italics indicate periods and
  masses.  
\label{fig:mobile}  }
\end{figure}

% HD 9500  19771  152027  156274

The   five   multiple  systems   studied   here   are  introduced   in
Table~\ref{tab:list}.  It gives the Washington Double Star (WDS) codes
\citep{WDS}, HD and Hipparcos  numbers, combined visual magnitudes and
parallaxes.  The  Gaia \citep{Gaia} parallaxes  of unresolved multiple
stars  are   notoriously  unreliable,   often  being  biased   by  the
photo-center  motion.  We  use unbiased  parallaxes of  distant single
components of  these systems where available  or, otherwise, dynamical
parallaxes derived from visual  orbits and estimated masses.  The last
four columns of Table~\ref{tab:list} give the types and periods of the
outer and  inner subsystems using  codes adopted in the  Multiple Star
Catalog \citep[MSC][]{MSC}: V -- visual orbit, A -- astrometric orbit,
and S1 -- single-lined spectroscopic  orbit.  All stars are located within
67\,pc from the Sun and have masses below 1.1 \msun. 
  Figure~\ref{fig:mobile} illustrates the structure of these systems
  and their parameters. Two quadruples, HD~12376 and HD~152027, have a
  3+1 hierarchy where a close inner binary has a tertiary 
  component, and this triple is orbited by the fourth,  more distant
  component. 

The data and method of orbit calculation are presented in
Section~\ref{sec:data}. Section~\ref{sec:systems} contains the
results for each of the hierarchies. The paper closes with a short
summary in section~\ref{sec:sum}. 

%---------------------------------------------------------
\section{Data and Methods}
\label{sec:data}

%------------------------------------
\subsection{Radial Velocities}
\label{sec:RV}

Radial velocities (RVs) of  these multiple systems have been monitored
as part  of the survey  of nearby solar-type stars.   The observations
have been obtained at  the Harvard-Smithsonian Center for Astrophysics
(CfA) using several instruments  on different telescopes.  The Digital
Speedometers \citep[see][]{Latham:1992} were  used on 1.5 m telescopes
at  the Oak Ridge  Observatory (Harvard,  Massachusetts) and  the Fred
L.\ Whipple  Observatory (Mount Hopkins,  Arizona), as well as  on the
Multiple Mirror Telescope  with an equivalent aperture of  4.5 m (also
on  Mount  Hopkins)  before  its  conversion  to  a  monolithic-mirror
telescope.  In these  instruments, intensified photon-counting Reticon
detectors delivered  a single echelle  order 45~\AA\ wide  centered at
5187~\AA\ (featuring  the \ion{Mg}{1}~b triplet) at  a resolving power
of 35,000.

We   also  used   the  Tillinghast   Reflector   Echelle  Spectrograph
\citep[TRES;][]{Szentgyorgyi:2007, Furesz:2008} attached  to the 1.5 m
telescope  on  Mount  Hopkins,   which  covers  the  wavelength  range
3900--9100~\AA\ in 51 orders at a resolving power of 44,000. 

RVs  were measured  by cross-correlation  using suitable  synthetic or
observed templates centered on the \ion{Mg}{1}~b triplet.  The CfA RVs
are given on the native  instrument system (a correction of +0.14 \kms
would be  needed to bring them  to the IAU system).   The RVs measured
with TRES have been corrected to the same zero point as the velocities
from the CfA  Digital Speedometers, so they also  need to be corrected
by +0.14 \kms to put them on the IAU system.

The orbits of HD~12376 and HD~89795 also use RVs measured with the CORAVEL
instruments located at the Haute Provence and La Silla observatories,
respectively. These observations were obtained in the framework of the
large survey of solar-type stars conducted in collaboration with the CfA team
by \citet{N04}; section 3.2.1 of their paper contains references on CORAVELs.   

The measured RVs  refer to the brightest star  dominating the combined
(blended)  spectra   of    multiple  systems.   Blending  with  other
components may reduce the amplitude of the RV variation.  This bias is
taken into consideration in the analysis of individual systems.

%------------------------------------
\subsection{Position Measurements}
\label{sec:pos}

All objects  except the  interferometric pair HD~190412  are classical
visual binaries.   Their orbits are based,  to a large  extent, on the
micrometer   measurements  made  by   various  observers   during  two
centuries. These  data were obtained  from the 
WDS database \citep{WDS} on our request. Obsolete discoverer's codes
are used in  the WDS to identify double stars,  e.g.  the pairs A~1813
and  BU~25 discoreved  by R.~Aitken  and S.~Burnham,  respectively. We
follow  the WDS  scheme and  designate  components by  one or  several
characters;  systems are denoted  by joining  their components  with a
comma.  For  example, in  a visual binary  A,B the inner  subsystem in
component A is  designated as Aa,Ab. This scheme  and its extension to
code the hierarchy is further explained in \citet{MSC}.

The subjective nature of visual measurements and different
qualification of observers do not allow consistent estimates of the
measurement errors. Some measurements, even those made by
well-recognized observers, turn out to be erroneous or misleading when
confronted with the full data set. In fitting the orbits, strong
outliers are rejected, and the remaining visual measurements are
assigned errors from 50 to 250 mas,  based on their scatter. 

Much   more  accurate   speckle-interferometric   measurements  became
available  starting  from the  late  1970s.  Typical  errors of  early
measurements on the 4 m class telescopes are about 5 mas. Measurements
made   with    smaller   telescopes   have    correspondingly   larger
errors. Modern speckle  interferometers based on solid-state detectors
reach     an     even      higher     accuracy     of     1--2     mas
\citep[e.g.][]{SOAR}.  Relative   positions  of  binaries   have  been
measured by  Hipparcos with  a typical accuracy  of 10 mas;  Gaia also
measured component's positions in some pairs wider than 0\farcs7.

%------------------------------------
\subsection{Orbit Calculation}
\label{sec:orb}

\begin{deluxetable*}{l  rrrr rrr r rr }
\tabletypesize{\scriptsize}
%\tablenum{7}
\tablewidth{0pt}
\tablecaption{Orbital Elements \label{tab:orb}}
\tablehead{
\colhead{WDS/system} &
\colhead{$P$} & 
\colhead{$T  $} &
\colhead{$e$} & 
\colhead{$a$} & 
\colhead{$\Omega$} &
\colhead{$\omega$} &
\colhead{$i$}  &
\colhead{$f$}  & 
\colhead{$K_1$} &
\colhead{$\gamma$}
 \\
\colhead{Name} &   
\colhead{(yr)} & 
\colhead{(yr)} &
 & 
\colhead{($''$)} & 
\colhead{(\degr)} &
\colhead{(\degr)} &
\colhead{(\degr)} & 
&
\colhead{(\kms)} &
\colhead{(\kms)} 
}
\startdata
02022+3643 AB,C     &  345    &  2254.7   & 0.533    &  1.450  & 317.3  & 319.5    & 138.9   & \ldots & \ldots  & 13.76 \\
\phn HD 12376     & fixed   &$\pm$1.8      &$\pm$0.018 &fixed &$\pm$3.9   &$\pm$3.3  &$\pm$1.5 & \ldots & \ldots  &$\pm$0.10 \\
02022+3643 A,B      & 12.903  & 1989.090    &0.415      & 0.1496   & 191.6      &  113.7 & 66.1    & 0.56     & 11.02    &  \ldots \\
\phn HD 12376     &$\pm$0.019 &$\pm$0.036 &$\pm$0.008  &$\pm$0.0017&$\pm$0.8  &$\pm$1.0 &$\pm$0.7 &$\pm$0.03 &$\pm$0.14 & \ldots \\
\hline
03122+3713 A,B      & 1500    & 1230.7  & 0.375 & 3.70     & 109.8     & 119.4    & 100.4    & \ldots   & \ldots   & \ldots \\
 \phn HD 19771    &  fixed  &$\pm$300 &$\pm$0.262  &fixed &$\pm$5.7&$\pm$39  &$\pm$1.8 & \ldots   & \ldots   & \ldots \\
03122+3713 Aa,Ab    & 1.13545 & 2002.844 &0.462    & \ldots & \ldots & 272.1   & \ldots  & \ldots   & 11.10    &$-$20.10 \\    
 \phn HD 19771   &$\pm$0.00006 &$\pm$0.003 &$\pm$0.006 &  \ldots & \ldots &$\pm$1.4 & \ldots  & \ldots   & $\pm$0.11   &$\pm$0.09 \\
\hline
10217$-$0946  A,B      & 475    & 2152.8  & 0.409 & 1.28     & 250.6     & 262.3    & 170.0    & \ldots   & 0.1   & $-$2.82 \\
 \phn HD 89795    &  fixed  &$\pm$2.0 &$\pm$0.008  &fixed &$\pm$16.1&$\pm$16.6  & fixed & \ldots   & fixed   & $\pm$0.06 \\
10217$-$0946  Aa,Ab    & 8.39 & 2013.85 &0.440    & 0.0734     & 208.2 & 139.3   & 160  & 0.193   & 1.235    & \ldots \\    
 \phn HD 89795   &$\pm$0.08 &$\pm$0.14 &$\pm$0.044 & fixed & $\pm$8.0 &$\pm$9.4 & fixed  & $\pm$0.009   & $\pm$0.088 & \ldots \\
\hline
16446+7145 Aa,Ab       & 39.72      & 1995.56    & 0.399 & 0.233        & 293.4   & 45.2    & 124.7   & \ldots   & 1.68    & $-$1.53  \\
 \phn HD 152027     &$\pm$0.78  &$\pm$0.33  &$\pm$0.017 &$\pm$0.004 &$\pm$1.3 &$\pm$2.5 &$\pm$1.5 & \ldots &$\pm$0.15 & $\pm$0.09 \\
16446+7145 Aa1,Aa2     & 1.1099   & 2001.144 & 0.115     & \ldots    & \ldots  & \ldots  & 73.9    & \ldots &   3.392 & \ldots \\
 \phn HD 152027 &$\pm$0.0004&$\pm$0.016 &$\pm$0.012 & \ldots    & \ldots  & \ldots  &$\pm$5.1 & \ldots &$\pm$0.048 & \ldots \\
\hline 
20048+0109 A,B       & 7.446     & 2018.701  & 0.202     & 0.150     & 247.7   & 239.3   & 32.3    & \ldots & 3.01  & $-$55.34 \\
 \phn HD 190412  &$\pm$0.025 &$\pm$0.034&$\pm$0.009 &$\pm$0.003 &$\pm$2.8 &$\pm$3.0 &$\pm$2.0 & \ldots &$\pm$0.10 &$\pm$0.03 \\
20048+0109 Aa,Ab     &0.68822    & 1998.714 & 0.039     & 0.026     & 239.5   & 267.1   & 31      & 0.35   & 6.167 & \ldots \\  
\phn HD 190412  &$\pm$0.00011&$\pm$0.030&$\pm$0.009& fixed     &$\pm$9.6 &$\pm$15.8& fixed   &$\pm$0.07 &$\pm$0.11 & \ldots
\enddata
\end{deluxetable*}

The orbital elements  derived here and their formal  errors are listed
in  Table~\ref{tab:orb},  in   common  notation.  The  ascending  node
$\omega$  refers to  the  main component  to  match the  RVs, and  the
longitude of periastron $\Omega$ is chosen accordingly to describe the
motion of the  secondary, in agreement with the  convention for visual
orbits.  In this way,  both positional measurements and RVs correspond
to the common set of elements. When no RV information is available,
both $\omega$  and  $\Omega$ can be simultaneously changed by 180\degr
~because the choice of correct orbit node remains
ambiguous. Consequently, the mutual inclination $\Phi$ can take two
alternative values.

The method  of orbit  calculation is described  in our  previous paper
\citep{TL2017}.\footnote{Codebase:
  \url{http://dx.doi.org/10.5281/zenodo.321854}} We fit simultaneously
elements  of the  inner and  outer orbits  to the  data  -- positional
measurements  and RVs.   Only one  inner  subsystem (in  HD 12376)  is
resolved.   However, in  two other  systems (HD  89795 and  190412) we
could detect  the wobble in the  motion of the outer  binary caused by
the  subsystems and  determine  their astrometric  orbits. The  wobble
factor $f$ is the ratio of the astrometric and full semimajor axes. It
depends on the mass ratio $q$ and light ratio $r$ in the subsystem: $f
= q/(1+q)  - r/(1+r)$. When  relative positions refer to  the resolved
inner  pair or  when the  inner companion  is faint  ($r \ll  1$), the
second term is irrelevant and $f$ becomes directly related to $q$. The
inner semimajor axis of subsystems that are not directly resolved is
calculated from the period and the estimated mass sum by  Kepler's Third law. 

In fitting the orbits, weights inversely proportional to the
squares of the measurement errors are used. An orbit should have 
$\chi^2/N \sim 1$ when  residuals match the errors. As noted above,
the errors of positional measurements are not known in advance; they
are assigned based on the measurement technique and, if necessary,
adjusted to reduce the impact of outliers. In contrast, the RV errors
are known. The condition $\chi^2/N \sim 1$ should be satisfied for
both RVs and position measurements to balance their relative influence
on the solution. 

Long  periods  of  some outer  orbits  exceed  the  time span  of  
observations. Short  observed arcs do not fully  constrain these orbits
and, instead, match a  family of different orbits.  For  such long-period pairs,
we  provide a  representative member  of  this family  by fixing  some
elements  and fitting  the remaining elements.   In doing  so, we  match  the poorly
constrained  orbit to  the expected  mass sum.   These  notional outer
orbits are  needed here as a  reference to measure  wobble. They
also constrain, to some extent, the mutual inclination. Exploration of
all potential  long-period orbits  compatible with the  observed 
arcs is not deemed to be necessary here.

The RVs used in spectroscopic  and combined orbits and their residuals
are    listed    in     Table~\ref{tab:rv},    published    in    full
electronically. Its  first two columns  contain the HD number  and the
component to which the RV refers, followed by the Julian date, RV, its
error,  and residual  to  the  orbit. The  last  column indicates  the
instrument.   Individual positional  measurements and  their residuals
from the orbits are  listed in Table~\ref{tab:speckle}, also published
electronically. Its  second column indicates  the pair (note  that for
HD~12376, A,C refers  to measurements that resolve the  inner pair and
AB,C  to the unresolved  measurements between  photo-center of  AB and
C). Then  follow the data, position angle,  separation, adopted error,
and residuals in  angle and separation. The last  column specifies the
techniques  used.  For the  sake of  completeness, elements  of visual
orbits found in the literature are reproduced in Table~\ref{tab:old}.

\begin{deluxetable*}{l  rrrr rrr l }
\tabletypesize{\scriptsize}
%\tablenum{7}
\tablewidth{0pt}
\tablecaption{Previously Published Orbital Elements \label{tab:old}}
\tablehead{
\colhead{WDS/system} &
\colhead{$P$} & 
\colhead{$T  $} &
\colhead{$e$} & 
\colhead{$a$} & 
\colhead{$\Omega$} &
\colhead{$\omega$} &
\colhead{$i$}  &
\colhead{Reference}
 \\
 &   
\colhead{(yr)} & 
\colhead{(yr)} &
 & 
\colhead{($''$)} & 
\colhead{(\degr)} &
\colhead{(\degr)} &
\colhead{(\degr)} &
}
\startdata
02022+3643 AB,C     &  310    &  2210.24   & 0.636    &  1.790  & 168.5  & 135.5    & 113.2   & \citet{Nov2006e} \\
02022+3643 A,B      & 12.94  & 1989.06    &0.404      & 0.15   & 191.4   &  295.1 & 67.0    & \citet{Hrt2000a} \\
03122+3713 A,B      & 871    & 1454  & 0.710 & 3.66     & 111.0     & 105.0    & 98.0   & \citet{WSI2004a} \\
10217$-$0946  A,B      & 817    & 2295  & 0.0 & 1.837     & 0.2     & 0.0    & 141.2    & \citet{Zir2012b} \\
16446+7145 Aa,Ab       & 62.0  & 1977.5  & 0.19 & 0.263   & 64.7  & 358.0  & 121.0   & \citet{Hei1997}  \\
20048+0109 A,B       & 7.83   & 2018.61  & 0.256     & 0.164     & 74.3   & 47.6   & 40.5    & \citet{Tok2018i} 
\enddata
\end{deluxetable*}

%\input{rv_small.tex}
%\input{speckle_small.tex}

% radial velocities
\begin{deluxetable*}{r l c rrr l }    
\tabletypesize{\scriptsize}     
\tablecaption{Radial Velocities and Residuals (fragment)
\label{tab:rv}          }
\tablewidth{0pt}                                   
\tablehead{                                                                     
\colhead{HD} & 
\colhead{Comp.} & 
\colhead{JD} & 
\colhead{RV} & 
\colhead{$\sigma$} & 
\colhead{(O$-$C)$$ } &
\colhead{Instr.\tablenotemark{a}}
 \\
 & & 
\colhead{(JD $-$24,00,000)} &
\multicolumn{3}{c}{(km s$^{-1}$)}  &
}
\startdata
 12376 & A & 43519.2870 &      2.830 &      1.000 &      0.110 & COR \\
 12376 & A & 43519.2910 &      4.160 &      1.000 &      1.440 & COR \\
 19771 & Aa & 49558.9932 &    $-$26.610 &      0.490 &  $-$0.131 & CfA 
\enddata 
\tablenotetext{a}{Instruments:
CfA: CfA digital speedometers;
MMT: CfA speedometer on MMT;
TRES: TRES;
COR: CORAVEL;
TS2002: \citet{TS2002};
Butler: \citet{Butler2017} with a $-$55.04 \kms offset.
}
\end{deluxetable*}

% radial velocities
\begin{deluxetable*}{r l l rrr rr l}    
\tabletypesize{\scriptsize}     
\tablecaption{Position Measurements and Residuals (fragment)
\label{tab:speckle}          }
\tablewidth{0pt}                                   
\tablehead{                                                                     
\colhead{HD} & 
\colhead{Syst.} & 
\colhead{Date} & 
\colhead{$\theta$} & 
\colhead{$\rho$} & 
\colhead{$\sigma_\rho$} & 
\colhead{(O$-$C)$_\theta$ } & 
\colhead{(O$-$C)$_\rho$ } &
\colhead{Ref.\tablenotemark{a}} \\
 & & 
\colhead{(JY)} &
\colhead{(\degr)} &
\colhead{(\arcsec)} &
\colhead{(\arcsec)} &
\colhead{(\degr)} &
\colhead{(\arcsec)} &
}
\startdata
12376 & A,B &  1908.8100 &    188.5 &   0.2000 &   0.0500 &      2.2 &   0.0396 & M  \\
12376 & A,B &  1918.6600 &    163.0 &   0.2100 &   0.1000 &      6.1 &   0.0800 & M \\
12376 & A,C  &  1999.8843 &    207.6 &   1.4940 &   0.0050 &     $-$0.2 &  $-$0.0025 & s  \\
12376 & A,C &  2010.7185 &    200.0 &   1.6063 &   0.0050 &     $-$0.1 &  $-$0.0002 & s \\
12376 & AB,C  &  1991.2500 &    216.0 &   1.4210 &   0.0200 &      1.1 &   0.1030 & H  \\
12376 & AB,C  &  2015.5000 &    199.2 &   1.5520 &   0.0050 &      0.2 &  $-$0.0031 & G 
\enddata 
%\tablerefs{
\tablenotetext{a}{
C: CCD measurement; 
G: Gaia;
H: Hipparcos;
M: visual micrometer measurement;
P: photographic measurement;
S: speckle interferometry at SOAR;
s: speckle interferometry at other telescopes.
}
\end{deluxetable*}

%------------------------------------
\subsection{Gaia Astrometry and Photometry}
\label{sec:gaia}

The 5-parameter astrometric solutions in the second Gaia data
release \citep{Gaia} do not account for non-linear motion of the
photo-center. This affects the measured parallaxes and proper motions
(PMs) of unresolved binaries. Unusually large errors of the Gaia astrometry
help to distinguish such cases, but the actual bias can exceed even
these inflated errors. For this reason the Gaia PMs of unresolved
binaries may not accurately reflect the PM of their photo-centers. 

The $V$ magnitudes  of individual components resolved by Gaia are calculated from the
$G$ magnitudes and the BP$-$GP colors using  the prescription  given on  its web
site.\footnote{See Chapter  5.3.7 of  Gaia DR2  documentation at
  \url{https://gea.esac.esa.int/archive/documentation/GDR2/}.   }
Effective temperatures provided by Gaia help to constrain spectral
types of components. 
Masses  in Table~\ref{tab:mass}  are estimated from  the absolute visual
magnitudes  using  the standard  relations  of \citet{Pecaut2013}  for
main-sequence stars. Masses  of spectroscopic secondary components are
estimated from  their RV amplitudes  or wobble, as explained below for each
system.

\begin{deluxetable}{r c c  c }
\tabletypesize{\scriptsize}
%\tablenum{7}
\tablewidth{0pt}
\tablecaption{Visual Magnitudes and Masses \label{tab:mass}}
\tablehead{
\colhead{HD} &
\colhead{Component} & 
\colhead{$V$} &
\colhead{Mass}  \\
&  &
\colhead{(mag)} & 
\colhead{(\msun)}
}
\startdata
12376 & A  & 8.80 & 0.95 \\
      & B   & 9.00 & 1.62 \\
      & C  & 10.79 & 0.72 \\
19771 & Aa & 7.88  & 1.08 \\
      & Ab & \ldots & 0.47 \\
      & B  & 8.16  & 1.03 \\
89795  & Aa & 8.44 & 1.07 \\
        & Ab & \ldots & 0.27 \\
       & B & 9.12 & 0.96 \\
152027  &Aa1 & 9.15 & 0.98  \\
        & Aa2 & \ldots & 0.6: \\
        & Ab & 10.21 & 0.83 \\
        & B   & 13.88 & 0.54 \\
190412  & Aa  & 7.75  & 0.97 \\
        & Ab  & \ldots & 0.45 \\
        & B   & 11.37 & 0.61 
\enddata
\end{deluxetable}

%---------------------------------------------------------
\section{Individual Systems}
\label{sec:systems}

\subsection{02022+3643 (HD 12376)}
\label{sec:12376}

\begin{figure}
\epsscale{1.1}
%\plotone{HD12376TESS.eps}
\plotone{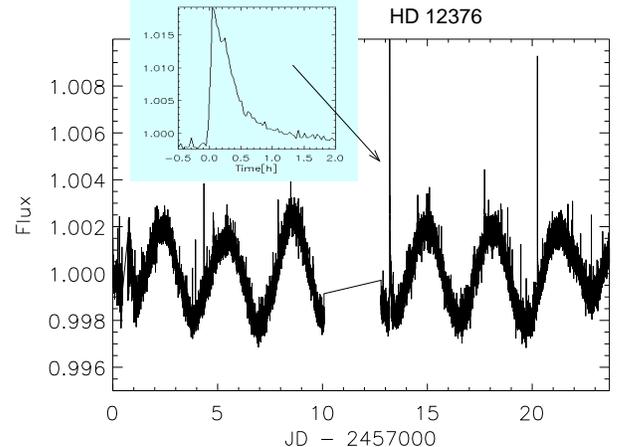}
\caption{Flux modulation of HD 12376 by starspots and flares as recorded
  by TESS. The insert shows one flare plotted on a different time scale. 
\label{fig:12376flux}  }
\end{figure}

\begin{figure}
\epsscale{1.0}
%\plotone{HD12376in.eps}
\plotone{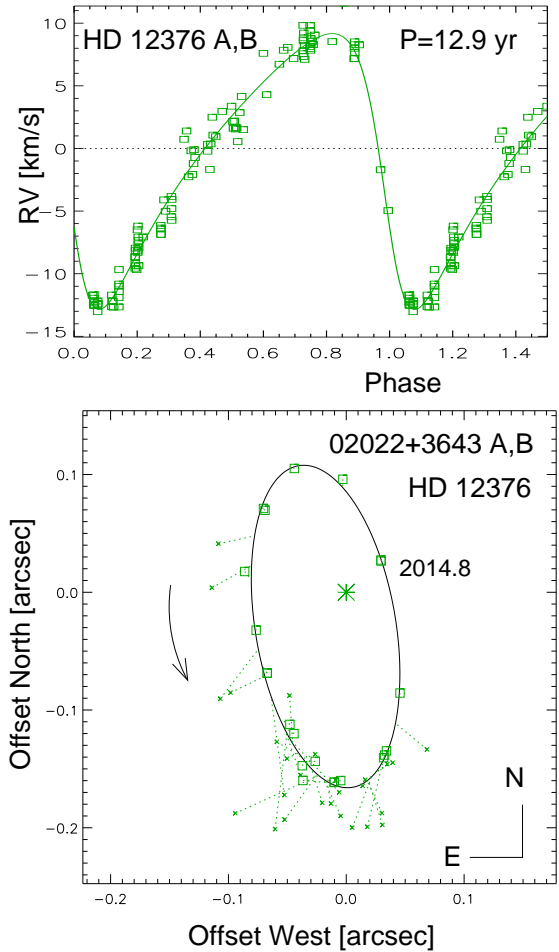}
\caption{Orbit of HD 12376 A,B with $P=12.9$ yr. The upper panel plots
  RVs  vs.   orbital  phase,  the  lower  panel   plots  the  position
  measurements (squares  for speckle  data, crosses for  less accurate
  micrometer data). Dotted lines connect the measured positions to the
  ephemeris positions on the orbit (ellipse).
\label{fig:12376in}  }
\end{figure}

\begin{figure}
\epsscale{1.0}
%\plotone{HD12376out.eps}
\plotone{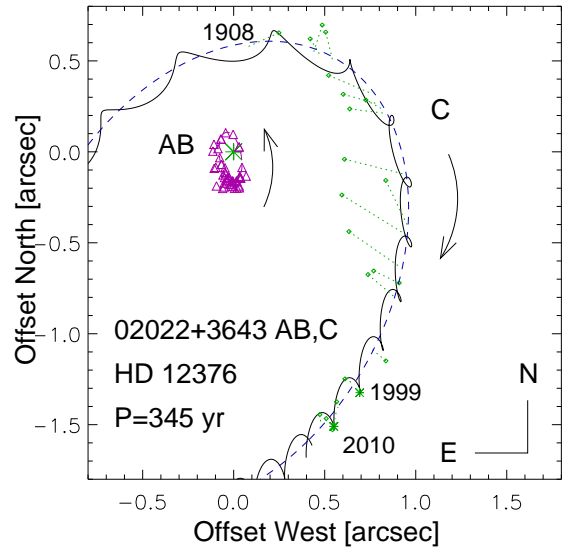}
\caption{Fragment of the orbit of HD 12376 AB,C with $P=345$ yr. Solid
  line and asterisks depict the motion of A,C (i.e. the resolved inner
  system), while dashed line and crosses refer to the unresolved orbit
  of AB,C. The  inner orbit of A,B  is shown for reference on the same  scale by the
  magenta ellipse and triangles. Note opposite rotation
  sense of the inner and outer orbits.
\label{fig:12376out}  }
\end{figure}

This  is  a   quadruple  system  of  3+1  hierarchy,   also  known  as
ADS~1613. The visual triple A~1813 A,B and AB,C has been discovered by
R.~Aitken in  1908.  Both  inner and outer  systems have  known visual
orbits, refined  here. Moreover, the inner secondary component
B is  a double-lined spectroscopic  pair with $P=3.082$ d  (M.  Mayor, 1995,
private   communication).   Precise photometry   by the
Transiting  Exoplanet  Survey  Satellite  \citep[TESS,][]{TESS}  shows
flux modulation with a  period of 3.1 d, presumably  caused by starspots on
tidally  locked   stars  Ba  and   Bb,  and  flares  in   their  active
chromospheres        (Figure~\ref{fig:12376flux}).\footnote{A photometric
  period  of   1\fd4937  was  previously  determined   for  this  star
  designated as  V371~And. This can be  an alias  of the  3 day period.}
The  close  binary Ba,Bb  is  not  eclipsing.   The Gaia  parallax  of
19.44$\pm$0.17 mas may  be biased, and we adopt  instead the dynamical
parallax of  19.9 mas  deduced from the  well-defined inner  orbit and
the estimated mass sum.

The RVs refer to the blended light  of A and B (the contribution of C,
at 1\farcs5 from  AB and 2.7 mag fainter, can  be neglected). Only the
lines of the brightest star  A are measured.  Broad lines produced by
Ba and Bb are detectable in the cross-correlation function, but no RVs
were extracted so far.

The inner combined visual/spectroscopic orbit with $P=12.9$ yr is very
well  constrained by  the century-long  astrometry  and the   RVs
 (Figure~\ref{fig:12376in}).   The  RV  residuals of  1  \kms
exceed the internal  RV errors, possibly because of  blending with the
lines of Ba and Bb.  We adopted  RV errors of 1 \kms and discarded two
most discrepant measurements.  The weighted residuals of speckle positions are
6\,mas. The orbital inclination, the  estimated mass of A, 0.95 \msun,
and its RV amplitude correspond to  the mass of 1.62 \msun for B. This
matches the standard relations if B  is composed of two equal stars of
$V=9.75$ mag each. The mass sum  of AB, 2.57 \msun, corresponds to the
dynamical parallax of 19.9\,mas.

Incomplete  coverage of  the  outer orbit  (Figure~\ref{fig:12376out})
leaves its elements less well  defined. After the initial free fit, we
fixed the outer  period and semimajor axis to  match the expected mass
sum of  3.3 \msun  and the dynamical  parallax deduced from  the inner
orbit.  Most available measurements refer to the unresolved outer pair
AB,C,  but  two  resolved   measurements  of  A,C  were  published  by
\citet{Hor2017}.  They allow us to determine the wobble factor $f=0.56
\pm 0.03$. Obviously,  B is more massive than  A; our estimated masses
imply $f=0.63$.

The outer orbit corresponds to the RV amplitude of 1.5 \kms for AB and
predicts   a    negative   RV   trend   tentatively    seen   in   the
residuals. However,  the 18-yr RV coverage  (1978--1996) is  not long
enough  to measure the  outer RV  amplitude. The  choice of  the outer
nodes  remains   ambiguous.  It   corresponds  to  the   mutual  orbit
inclination of $\Phi  = 131 \pm 3$\degr. The inner  and outer pairs of
ADS~1613          rotate           in          opposite          sense
(Figure~\ref{fig:12376out}).  Alternative  choice  of  the  outer  node
corresponds  to orthogonal  orbits  ($\Phi =  87$\degr), which  would
produce  a strong modulation  of inner  eccentricity and  therefore seems
unlikely.

%------------------------------------
\subsection{03122+3713 (HD 19771)}
\label{sec:19771}

\begin{figure}
\epsscale{1.0}
%\plotone{HD19771in.eps}
\plotone{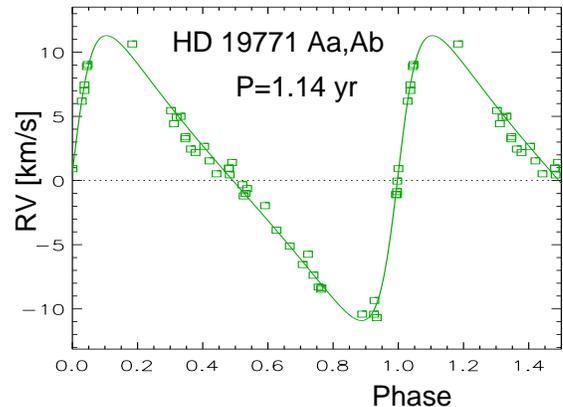}
\caption{Spectroscopic orbit of HD 19771 Aa,Ab.
\label{fig:19771in}  }
\end{figure}

\begin{figure}
\epsscale{1.0}
%\plotone{HD19771out.eps}
\plotone{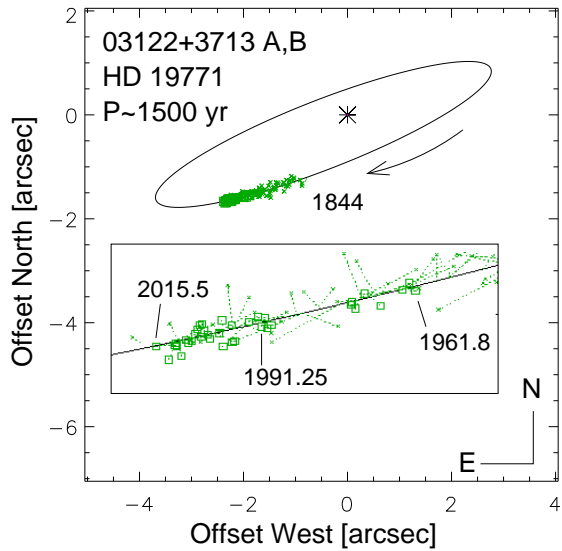}
\caption{Tentative visual orbit of WDS 13122+3713 A,B (HD 19771). Squares
  show positions with errors of less than 60 mas, crosses are less
  accurate micrometer measurements. The insert shows the latest data in detail.
\label{fig:19771out}  }
\end{figure}

The  outer  pair is  a  classical  visual  binary STF~360  (ADS  2390)
discovered  by W.~Struve in  1828 at  1\farcs2 separation.   It slowly
opened up and is now at 2\farcs9.  Only a small arc of the outer orbit
is  covered  and  its   long  period  remains  poorly  constrained;  a
preliminary orbit  with $P=871$ yr was  published by \citet{WSI2004a}.
RV  variability of  the brighter star A was  detected by  \citet{N04}; its
orbit with  a period of 1.14  yr is determined  here.
Motion of  the  photo-center caused  by the one-year  subsystem
seriously  biases the  Gaia  parallax, 24.42$\pm$0.66  mas.  The  more
reliable parallax  of star B,  20.58$\pm$0.09 mas, is adopted  here as
distance to  the system.  

There  was a suspicion  that RV  of B  is also  variable with  a small
amplitude.   However, close  proximity of  the two  stars  might cause
partial mixing of their light in the slit, hence the RV variability of
B remains  questionable. \citet{TS2002} published  30 RVs of A  and 23
RVs of  B monitored with a CORAVEL-type  correlation instrument during
six years. Their data do not confirm the RV variability of B.

The    spectroscopic    orbit     of    Aa,Ab    is    well    defined
(Figure~\ref{fig:19771in}). We  fitted the orbit separately  to 16 RVs
from CfA  covering the  period 1994-2014 or  27 RVs  from \citet{TS2002}
measured in 1992-1998. Both data  sets show excessive noise, while the
systemic velocities  agree within 0.1  \kms. The final orbit  uses all
data \citep[the  errors of  RVs from][are set  to 0.6  \kms]{TS2002} and
leaves the  weighted rms residuals of  0.43 \kms.  The  orbit of Aa,Ab
corresponds to the minimum mass of 0.47 \msun for Ab.  The mass can be
larger,  but not  by too  much,  otherwise the  lines of  Ab would  be
detectable in the spectrum.  

A plausible but otherwise arbitrary orbit of the outer visual pair A,B
with  $P=1500$  yr is  illustrated  in Figure~\ref{fig:19771out}.  Its
period and semimajor axis are chosen  to match the expected mass sum of
2.58  \msun.  Owing  to  the  wide separation,  there  are no  speckle
measurements.  Instead,  the most accurate astrometry  is furnished by
the  photographic  and  CCD  measurements,  as well  as  by  Gaia  and
Hipparcos.

Residuals to the orbit of A,B  were analyzed to search for a potential
wobble signal. No obvious periodicity was found. If only accurate data
(errors less than  60 mas) are considered and  the residuals exceeding
0\farcs1 are ignored, the rms scatter of the remaining residuals is 21
mas  in  separation  and  0\fdg25  in  angle  (12  mas  in  tangential
direction).   It is  well  known that  separations  are measured  less
accurately  than angles  because  of the  partial  image blending,  so
larger  residuals in  separation are  natural.  A  tentative  orbit of
Ba,Bb deduced from the CfA RVs has a period of 15.4 yr, semi-amplitude
1.75 \kms, and implies  a Ba,Bb semimajor axis of  0\farcs13.  The RVs
from \citet{TS2002} do not match this orbit, however.  Absence of wobble
larger than $\sim$20  mas constrains parameters of a  potential subsystem in the
component B. Its  existence cannot be ruled out  but, so far, remains
unproven.

The calculated amplitude  of the wobble caused by  the subsystem Aa,Ab
is about  8 mas. Such a weak  signal is not detectable  at the current
accuracy of  A,B measurements but  biases the Gaia parallax,  as noted
above.  Hopefully, future Gaia data releases will separate orbital and
parallactic motions of  A and will define its  astrometric orbit.  The
PM of B  relative to A measured by Gaia is  $(+3.0, +6.5)$ \masyr. The
relative motion speed in the outer orbit is $(+8.5, -2.0)$ \masyr. The
discrepancy is  presumably caused by  the motion of Aa,Ab  and matches
the PM anomaly \citep{Brandt2018}.

\subsection{10217$-$0946 (HD 89795)}
\label{sec:89795}

\begin{figure}
\epsscale{1.1}
%\plotone{BU25b.eps} 
\plotone{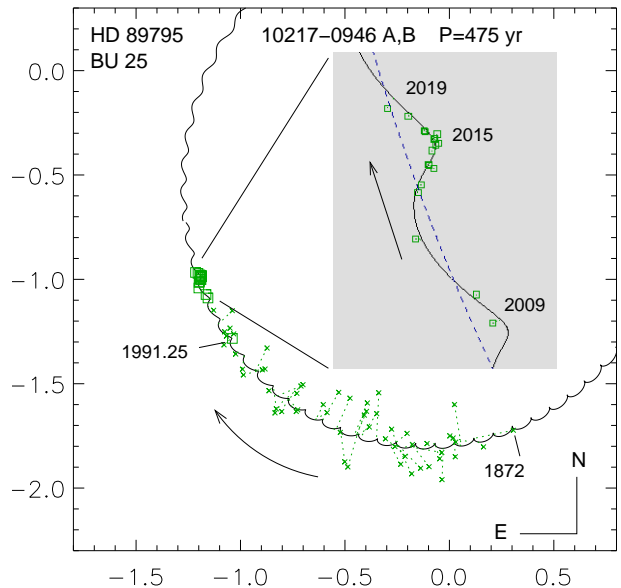} 
\caption{Orbital motion of HD 89795  (WDS J10217$-$0946, BU~25) A,B. Observed part of
  the  orbit (including  wobble) is  plotted  by the  full line.   The
  micrometer  and  low-accuracy speckle  measurements  are plotted  by
  crosses,  the   accurate  speckle  and   Hipparcos  measurements  by
  squares. The insert shows the orbit segment covered at SOAR.
\label{fig:89795out} }
\end{figure}
% BU25.fig

\begin{figure}
\epsscale{1.1}
%\plotone{HD89795in.eps} 
\plotone{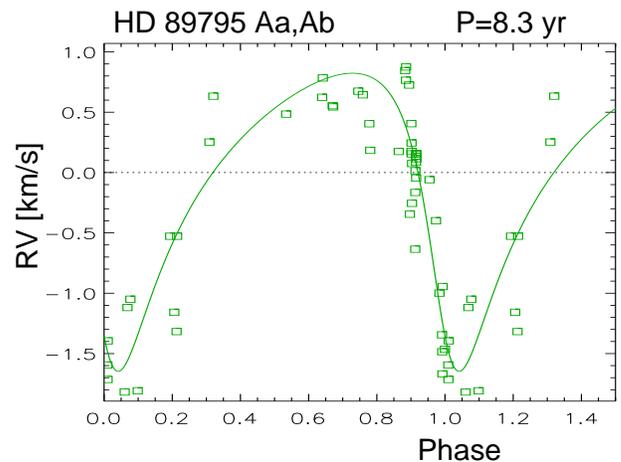} 
\caption{Spectroscopic orbit of the inner 8.3 yr subsystem HD 89795 Aa,Ab. 
\label{fig:89795in} 
}
\end{figure}
% BU25.fig

The outer  pair of HD  89795 (WDS J10217$-$0946, BU~25, ADS~7738) was
discovered by S.~W. Burnham in 1872  at 1\farcs75 separation. Since then, it
has turned  clockwise by 61\degr.  Gaia parallaxes 
are 16.25$\pm$0.11 mas for A and 15.88$\pm$0.08 mas for B. The latter,
unbiased by the inner subsystem, is adopted as true distance.

%The DR2 relative position is 1\farcs546 at 130\fdg15, with $\Delta G =
%0.61$ mag.

A spectroscopic subsystem in the component A was found by \citet{N04}.
They  observed  this  star   using  CORAVEL  from  1988.2  to  1997.2.
Follow-up   observations   were  conducted   with   the  CfA   digital
spectrometer from 2004.0  to 2009.3 and the last  RV was measured with
TRES in 2014.1.  The RV data cover 25.2 yr or  3.1 inner periods.  The
RVs of  the visual  secondary star B   were
also measured at CfA 15 times. Their mean value is $-$3.28 \kms with the rms
scatter of 0.69 \kms around the mean (un-weighted).

The preliminary  spectroscopic orbit  of Aa,Ab with  a period  of 3056
days (8.37 yr)  computed by D.~W.~L.  in 2009  implied an inner semimajor
axis of  76\,mas.  Therefore, a wobble  in the observed  motion of the
outer  pair  A,B was  expected.   This pair  has  been  placed on  the
speckle interferometry  program at  the 4.1  m  Southern Astrophysical
Research  (SOAR) telescope and  monitored for ten  years, from  2009.26 to
2019.95  \citep[][and  references  therein]{SOAR}.  The  spectroscopic
subsystem  has not  been resolved,  but accurate  measurements enabled
detection of the expected wobble.

A  preliminary  orbit   of  A,B  with  $P=817$  yr   was  computed  by
\citet{Zir2012b}, assuming zero  eccentricity.  The short observed arc
does not constrain the outer  orbit well enough.  Here elements of the
outer  and inner  orbits are  fitted  jointly to  determine the  inner
astrometric orbit.  As no resolved measurements of the inner subsystem
are available, its semimajor axis was fixed at the calculated value of
73.4\,mas. The resulting wobble factor $f=0.193 \pm 0.009$ is measured
reliably and  corresponds to the  wobble amplitude $\alpha =  14.2 \pm
0.7$ mas.

Finding an optimum set of orbital elements describing all data was not
trivial because  both inner and outer orbits  have small inclinations,
i.e. are  oriented almost in the  plane of the sky.  This follows from
the   small  RV   amplitude  of   the   inner  orbit   and  from   the
near-coincidence of the mean RVs of  A and B; their difference is much
less than a few \kms  implied by our first-guess outer circular orbit.
In the final iteration, we fixed the outer inclination to 175\degr ~to
obtain a small  but non-zero RV difference between A  and B. The final
values of $P$  and $a$ were also  fixed to match the mass  sum of 2.33
\msun. The  RV difference between  A and B  is so small that  even its
sign  (which defines  the  correct node  of  the outer  orbit) is  not
established  reliably.  No  RV  trend   due  to  the  outer  orbit  is
detectable.  So,  we simply fixed the  outer RV amplitudes  to 0.1 and
0.15  \kms.   These  amplitudes  have  only  a  minor  effect  on  the
center-of-mass velocity.

The orbital solution including wobble  results in the rms RV residuals
of  0.37 \kms,  matching the  RV errors  ($\chi^2/N \approx  1$).  The
weighted rms astrometric residuals are  3 mas.  They are comparable to
the residuals  of the calibration  binaries observed at SOAR  to their
modeled motion \citep{SOAR}.  The Gaia measurement of A,B was not used
to avoid  a small difference  in the scale  noted before (it  fits the
orbit well if the separation is increased by 9 mas).  Residuals to the
orbit    fitted    with    zero    wobble   increase    to    14\,mas.
Figure~\ref{fig:89795out} shows the observed  arc of the A,B orbit and
the  zoom on  the SOAR  data.  Figure~\ref{fig:89795in}  gives  the RV
curve  of  the inner  orbit  with  the  outer orbit  subtracted.   The
elements   of   both   orbits   and   their  errors   are   given   in
Table~\ref{tab:orb}. The  preliminary spectroscopic orbit  of Aa,Ab by
D.~W.~L. had $K_1 = 1.23$  \kms, $\gamma = -2.90$ \kms, $e=0.50$, $\omega
= 140\fdg6$. Owing  to the small RV amplitude  caused by face-on orbit
orientation, positional  measurements have a  substantial influence on
the combined inner orbit.

The  masses  of stars  Aa  and B  estimated  from  their absolute  $V$
magnitude  using   standard  relations \citep{Pecaut2013}   are  1.07  and   0.96  \msun,
respectively.  The wobble  factor $ f = 0.193$  implies the inner mass
ratio $q_{\rm Aa,Ab}  = f/(1-f) = 0.25$, hence the mass  of Ab is 0.27
\msun.  The inner RV amplitude and inclination fixed to $i_{\rm Aa,Ab}
=  160\degr$ match  this mass.   A free  fit leads  to  $i_{\rm Aa,Ab}
\approx 180\degr$, incompatible with the observed RV variation of Aa.

Both  inner  and outer  binaries  have retrograde  (clockwise)
motion.  The adopted node of the outer orbit corresponds to the mutual
inclination $\Phi = 17 \pm 2$\degr; the alternative value obtained
by  swapping the  nodes is  24\degr. As  both orbits  are  seen almost
face-on, the two  values of  mutual inclination are close to
each other.

Gaia measured a notable difference between the PMs of B and A, $(-3.1,
+10.5)$ \masyr in R.A.  and declination, respectively.  Our two orbits
predict  the motion  of B  relative to  A in  2015.5 with  a  speed of
$(-4.2, +8.3)$  \masyr, in   reasonable agreement with Gaia.   
Our astrometric  orbit of Aa,Ab approximately matches  the observed PM
anomaly --- the difference between  the short-term PM measured by Gaia
and  the  long-term PM  deduced  from  Gaia  and Hipparcos  positions.
According to  \citet{Brandt2018}, it is $(8.26,  2.06)$\,\masyr in the
R.A.  and declination  directions.  The  inner orbit  predicts  the PM
anomaly  of  $(-8.9, -4.7)$  \masyr.  The  sign  is inverted  because,
according to  the convention,  visual elements  describe  motion
of Ab around Aa while the PM anomaly refers to Aa.

%------------------------------------
\subsection{16446+7145 (HD 152027)}
\label{sec:152027}

\begin{figure}
\epsscale{1.0}
%\plotone{HD152027in.eps}
\plotone{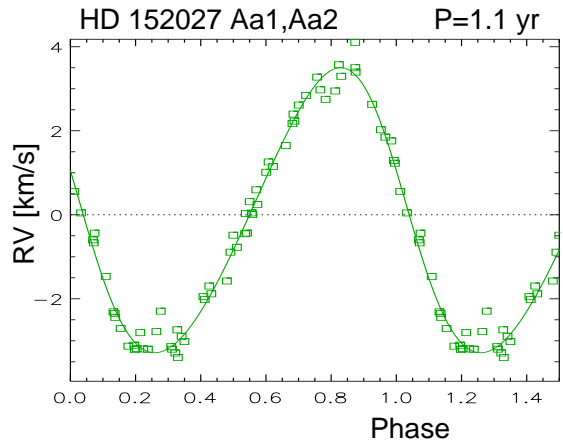}
\caption{Spectroscopic orbit of HD 152027  Aa1,Aa2.
\label{fig:152027in}  }
\end{figure}

\begin{figure}
\epsscale{1.0}
%\plotone{HD152027out.eps}
\plotone{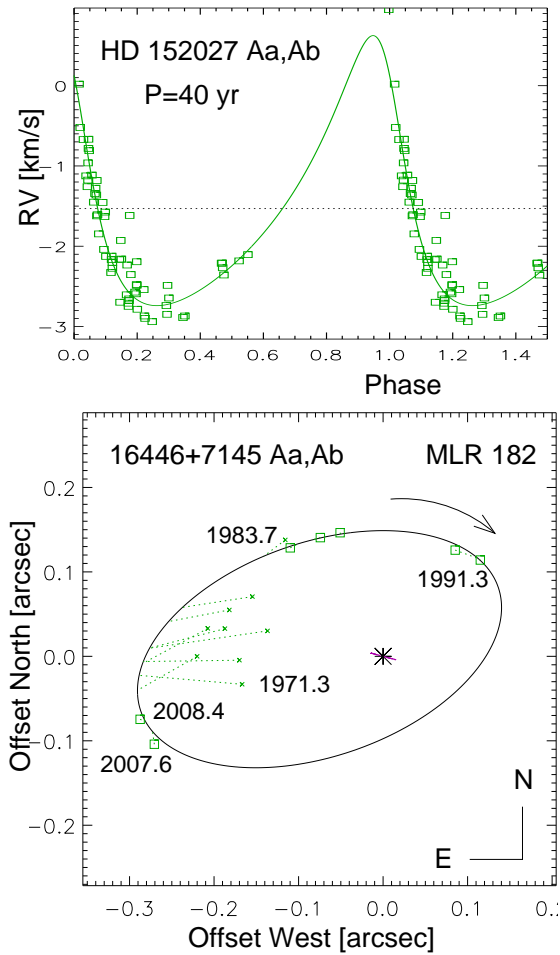}
\caption{Combined orbit of HD 152027 (WDS J16446+7145, MLR 182) Aa,Ab
\label{fig:152027out}  }
\end{figure}

This  is a  quadruple system  of  3+1 hierarchy.  The outer  27\farcs6
common proper motion pair UC~3219  A,B is physical. The parallax of the
faint ($V=13.88$ mag)  secondary star B, 14.854$\pm$0.035  mas, defines the
distance to the system (Gaia does not provide astrometry for the inner
triple). The intermediate subsystem  is a visual binary MLR~182 Aa,Ab
known since  1971. Finally, star  Aa is itself a spectroscopic  binary with a period
of 405 d discovered at CfA (Figure~\ref{fig:152027in}). It was
recognized as a spectroscopic triple and a preliminary outer period of
26.5 yr was determined by D.~W.~L. from RVs measured during a 22 yr period,
1995--2017. Seven RVs come from TRES. The weighted rms residuals to
the orbits determined here, 0.19 \kms, match the RV errors. However,
we ignored the first three slightly discrepant RVs.

The tentative visual orbit of  Aa,Ab with $P=63$ yr by \citet{Hei1997}
is revised here to $P=40$  yr using both the position measurements and
the RVs. The latest  two speckle measurements by \citet{Msn2011d}, made
in 2007  and 2008, at  first appeared highly discrepant.   B.~Mason has
checked them  on our request  and found no errors.   This confirmation
prompted us  to re-consider  the orbit and,  finally, a  good solution
that  fits both  RVs and  speckle data  was found  with the  change of
quadrants  in 2007  and 2008  (Figure~\ref{fig:152027out}).   One full
revolution is covered. However, the  new orbit deviates from the early
visual measurements that systematically under-estimate the separation.
The  highly  discrepant  visual  measurement  by Heintz  in  1996  was
rejected.   The new  orbit gives  the mass  sum of  2.44 \msun  for A.
Given the  small number of  accurate speckle measurements, we  did not
try to determine  the wobble.  The inner semimajor  axis is 18.5\,mas,
the  estimated  wobble  amplitude  is $\sim$7  mas. Frequent  speckle
measurements during a couple of years would easily detect the wobble,
while Gaia might provide an astrometric orbit in the future. 

The  visual  orbit  together  with  estimated  masses  implies  an  RV
amplitude of 3  \kms, whereas the measured amplitude  in the outer orbit,
 1.7 \kms,  corresponds to an unrealistically  small mass of star
Ab. A likely  explanation of this discrepancy is  blending of lines
that  reduces  the  measured  RV  amplitude. The  inner  amplitude  is
likewise reduced by both  blending and inclination, and we tentatively
assign a mass of 0.6 \msun to Aa2 to match approximately the measured
mass sum of the visual pair Aa,Ab.

%------------------------------------
\subsection{20048+0109 (HD 190412)}
\label{sec:190412}

\begin{figure}
\epsscale{1.0}
%\plotone{HD190412in.eps}
\plotone{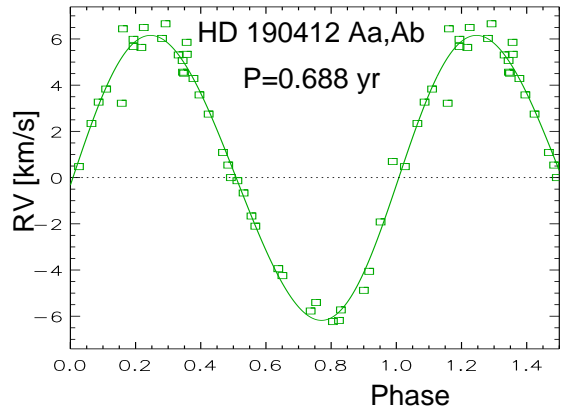}
\caption{Spectroscopic orbit of HD 190412 Aa,Ab.
\label{fig:190412in}  }
\end{figure}

\begin{figure}
\epsscale{1.0}
%\plotone{HD190412out.eps}
\plotone{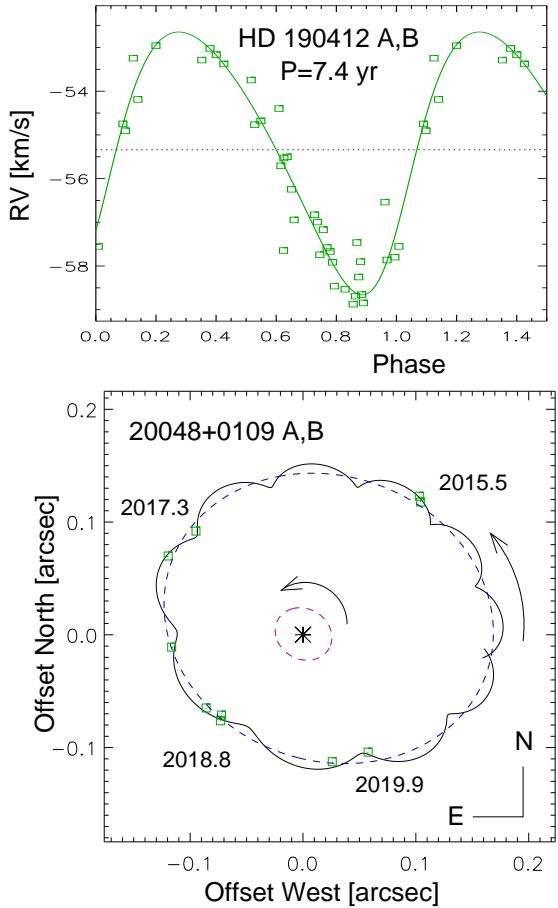}
\caption{Combined outer orbit of HD 190412 (TOK 699) A,B. The full
  line depicts orbit with wobble, the blue dashed line shows orbit
  without wobble, and the small ellipse at the center shows the inner
  orbit. 
\label{fig:190412out}  }
\end{figure}

This  is a nearby  (Gaia parallax  29.60$\pm$0.52 mas)  compact triple
system. It has been  identified as a single-lined spectroscopic triple
with periods  of 251 d  and 7.8 yr  in the CfA RV  survey; \citet{N04}
also found the RV to be variable. The outer pair produces acceleration
measured by Hipparcos.  It has been resolved by  the speckle camera at
SOAR in 2015 (TOK~699) and has been followed for 4.4 yr, covering most
of the outer orbit. Speckle photometry reveals a substantial magnitude
difference between  A and B,  2.5 mag  in $I$ and  3.9 mag in  $V$.  A
preliminary  7.8 yr  visual orbit  was published  by \citet{Tok2018i}.
Joint analysis  of the RVs  and speckle interferometry results  in the
combined outer orbit presented here.  Owing to the small period ratio,
the  orientation of  the  inner  orbit could  be  determined from  the
wobble.

Figure~\ref{fig:190412in} shows the inner spectroscopic orbit with the
contribution of the outer orbit subtracted.  We added five precise RVs
measured  by  \citet{Butler2017}  in  2003--2006  with  an  offset  of
$-$55.04 \kms deduced from our  preliminary orbits (these RVs are only
relative). Their  first RV  deviates by  1.5 \kms and  is given  a low
weight, as  well as  two CfA RVs.   The remaining data  are excellent,
with the  rms of only  0.14 \kms. Deviant  RVs are likely  produced by
blending    with    other   components.     Figure~\ref{fig:190412out}
illustrates the  outer orbit. Initially, the wobble  amplitude was set
to zero,  then the inner elements  $\Omega$, $i$, and  $f$ were fitted
jointly with both orbits.  The inner inclination is poorly constrained
by the data and was fixed  to 31\degr ~to match the inner RV amplitude
(a  free  fit  converges  to   $i=0$,  contradicting  the  RV
variability).  Including wobble in the model reduces the rms residuals
of  speckle measurements  from 6  mas to  2 mas,  indicating  that the
astrometric signal is significant. The  inner semimajor axis of 26 mas
is  computed and  the wobble  factor  $f=0.35\pm 0.07$  is found.   It
implies the inner mass ratio  of 0.54$\pm$0.15, while the RV amplitude
together  with   the  adopted  inner   inclination  implies  $q=0.46$.
Therefore, the masses of Aa, Ab,  and B of 0.97, 0.45, and 0.61 \msun,
respectively, approximately  match all  data.  The outer  RV amplitude
computed using  those masses is  3.32 \kms, the measured  amplitude is
3.01  \kms. Light  of  fainter stars  Ab  and B  slightly reduces  the
measured RV amplitudes by blending with the spectrum of Aa.

The outer orbit  is very well defined. Adopting the  outer mass sum of
2.04 \msun, we obtain the dynamical parallax of 31 mas, which compares
well   with   the  likely   biased   Gaia   parallax   of  29.6   mas.
\citet{Fuhrmann2019} took one high-resolution spectrum of HD~190412 in
2017.625 and  detected a  slight asymmetry of  the lines  
caused by  a weak component with an RV  shifted by +11.4 \kms relative
to the  stronger lines.  They modeled  the spectrum by  a combination of
three stars with effective temperatures  of 5650, 3900, and 4100 K and
solar metallicity. Our  orbits predict that on this date  the RV of B
was shifted  by +10.8  \kms relative to  Aa, in  qualitative agreement
with the Fuhrmann's result.

The most remarkable property of  this system is the orbit coplanarity:
the mutual inclination is $\Phi =  6 \pm 4$\degr, with a caveat that
the  inner  inclination is  not  well  constrained.   Note also the  small
eccentricities of  inner  and outer orbits.   The period  ratio is
10.82$\pm$0.04.   This   triple  system  belongs  to   the  family  of
``planeraty-like''  hierarchies like  HD 91962  \citep{Tok2015}, where
the inner periods of 0.47 and 8.8 yr are similar to those of HD 190412
and,  likewise,  the  orbits  have  moderate  eccentricities  and  are
approximately coplanar.  Similar architecture is  encountered in some other
multiple systems with low-mass components.

\section{Summary}
\label{sec:sum}

Our   work   contributed  inner   spectroscopic   orbits  in   several
hierarchical systems composed of  low-mass stars. In two instances (HD
89795  and 190412),  accurate speckle-interferometric  measurements of
the  outer pairs allowed  us to  detect wobble  produced by  the inner
subsystems. Astrometric orbits of  these subsystems are determined and
mutual inclinations  between inner  and outer orbits  are established;
both  systems are  approximately aligned  and co-rotate.  In contrast,
mutual inclination in the resolved triple HD~12376 is large, 131\degr;
the inner and outer  pairs are counter-rotating. Moreover, this system
contains an  inner close pair Ba,Bb. Architecture  of HD~12376 differs
markedly   from  that of planar   planetary-like   systems,  hinting   that
hierarchies in  the solar neighborhood might be  produced by different
mechanisms.

This study is  enabled by long-term monitoring of RVs  on a time scale
of  decades and position  measurements spanning  up to  two centuries.
The accuracy  and  resolution   of  modern  techniques,  especially  space
experiments like  Gaia, give access  to a vast number  of hierarchical
systems, but  the time  coverage is and  will remain a  major limiting
factor. Orbital periods of typical stellar and planetary systems range from
years to centuries and call for a patient accumulation of data.

\acknowledgements

D.~W.~L. thanks the many observers, especially Robert Stefanik, Joe Caruso, Joe Zajac, Mike Calkins, Perry Perlind, and Gil Esquerdo, who obtained observations with the CfA Digital Speedometers on telescopes in Massachusetts and Arizona, and with TRES on the 1.5-m Tillinghast Reflector at the Fred L. Whipple Observatory.
We thank Stephane Udry for sharing the radial velocities of HD 12376~ABC  from observations obtained with the northern CORAVEL on the SWISS 1.0-m telescope at the Observatoire de Haute Provence, and the radial velocities of HD 89795 from the southern CORAVEL on the Danish 1.54-m telescope at the European Southern Observatory on La Silla.

Some  data  used here  were  obtained  at  the Southern  Astrophysical
Research (SOAR) telescope.  This work used the SIMBAD service operated
by Centre des Donn\'ees Stellaires (Strasbourg, France), bibliographic
references from  the Astrophysics Data System  maintained by SAO/NASA,
and the Washington  Double Star Catalog maintained at  USNO.  We thank
B.~Mason for  extracting historic  measurements from the  WDS database
and  checking  his speckle  measurements  of  HD  152027.  This  paper
includes data collected  by the TESS mission funded  by NASA's Science
Mission directorate.
This  work has  made  use of  data from  the
European       Space       Agency       (ESA)       mission       Gaia
(\url{https://www.cosmos.esa.int/gaia}),  processed by  the  Gaia Data
Processing        and         Analysis        Consortium        (DPAC,
\url{https://www.cosmos.esa.int/web/gaia/dpac/consortium}).     Funding
for the DPAC has been provided by national institutions, in particular
the institutions participating in the Gaia Multilateral Agreement.

\facility {SOAR, Gaia, TESS, ADS, OHP:1.0 (CORAVEL), Danish 1.54-m Telescope (CORAVEL), FLWO:1.5m (TRES, DS), ORO:Wyeth (DS), MMT (DS)}

\end{document}